\documentclass[aps,prl,twocolumn,groupedaddress,showpacs]{revtex4}%
\usepackage{amsmath}
\usepackage{amsfonts}
\usepackage{amssymb}
\usepackage{graphicx}%

\begin{document}
\title{Quantum Metrology with Two-Mode Squeezed Vacuum:\\ 
Parity Detection Beats the Heisenberg Limit}
\author{Petr M.\ Anisimov}
\email{[]petr@lsu.edu}
\author{Gretchen M.\ Raterman}
\author{Aravind Chiruvelli}
\author{William N. Plick}
\author{Sean D. Huver}
\author{Hwang Lee}
\author{Jonathan P.\ Dowling}
\affiliation{Hearne Institute for Theoretical Physics and Department of Physics and Astronomy \\
Louisiana State University, Baton Rouge, LA 70803 }
\date{\today }

\begin{abstract}
We study the sensitivity and resolution of phase measurement in a Mach-Zehnder 
interferometer with two-mode squeezed vacuum ($\bar{n}$ photons on average). 
We show that super-resolution and sub-Heisenberg sensitivity is obtained 
with parity detection. In particular, in our setup, dependence of the signal 
on the phase evolves $\bar{n}$ times faster than in traditional schemes, and uncertainty 
in the phase estimation is better than $1/\bar{n}$.
\end{abstract}

\maketitle



Different physical mechanisms contribute to phase measurement. 
Thus, measuring phase provides insight into a number of physical 
processes.
Therefore, improved phase estimation benefits multiple 
areas of scientific 
research, such as quantum metrology, imaging, sensing, and 
information processing.
Consequently, enormous efforts have been devoted to improve the resolution and 
sensitivity of interferometers. Sensitivity is a measure of the uncertainty in 
the phase estimation, while resolution is rate at which signal changes with 
changing phase.

In what follows, we direct our attention to quantum interferometry. 
The benchmark that quantum interferometry is compared against is one 
with coherent 
light input and intensity difference measurement at the output of a 
Mach-Zehnder 
interferometer (MZI). 
In general, phase sensitivity of this benchmark is shot-noise limited, 
namely $\Delta \varphi=1/\sqrt{\bar{n}}$, where $\bar{n}$ is
the average number of photons. However, better sensitivity is
possible if nonlinear interaction between photons in the  MZI takes 
place \cite{Boixo09}.
In what follows, we only consider phase accumulation due to linear processes.

In 1981, Caves pointed out that by using coherent light and squeezed
vacuum one could beat the shot-noise limit $\Delta \varphi<1/\sqrt{\bar{n}}$ (super-sensitivity) 
\cite{ISI:A1981LL34600003}. 
In the work of Boto \emph{et al.}, it was shown that by exploiting quantum 
states of light, 
such as N00N states, it is possible to beat the Rayleigh diffraction limit 
in imaging and 
lithography (super resolution) while 
also beating the shot-noise limit in phase estimation
\cite{PhysRevLett.85.2733,PhysRevA.63.063407,ISI:000225565000028,ISI:000257343400003}. 
Finally, it was shown in Ref. \cite{giovannetti:010401} that input state 
entanglement is 
important in order to achieve super-sensitivity in a linear 
interferometer.

non-classical light
Experimental realization of these predictions have been hindered 
by the fact that
entangled states of light, with large numbers of photons, are 
difficult to obtain. 
Therefore we turn our attention to the brightest (experimentally available) 
non-classical light --- two-mode squeezed vacuum (TMSV). 
A state of TMSV is a superposition of twin Fock states $\left\vert \psi_{\bar{n}}\right\rangle=\sum
_{n=0}^{\infty}\sqrt{p_n\left(\bar{n}\right)}\left\vert n,n\right\rangle$, 
where the probability of a twin Fock state $\left\vert n,n\right\rangle =\left\vert n\right\rangle _{A}\left\vert
n\right\rangle _{B}$ to be present depends on
average number of photons in both modes of TMSV, $\bar{n}$, in the following way $p_n\left(\bar{n}\right)=(1-t_{\bar{n}})t^n_{\bar{n}}$
with $t_{\bar{n}}=1/\left(1+2/\bar{n}\right)$ \cite{gerrybook}.

\begin{figure}[ptb]
\includegraphics[width=6.9cm]{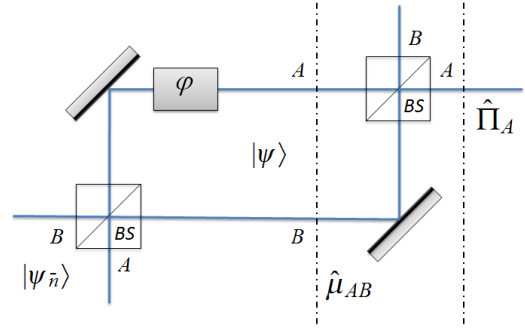}\caption{Mach-Zehnder interferometer
used in the calculations. Two-mode squeezed vacuum input state $\left\vert \psi_{\bar{n}}\right\rangle$ is indicated together with the intermediate state $\left\vert \psi\right\rangle =\hat{P}_{\varphi}\hat{U}\left\vert \psi_{\bar{n}}\right\rangle$. 
Vertical dash-dot lines indicate places where two measurements $\hat{\mu}_{AB}$ and $\hat{\Pi}_{A}$ are to be implemented.}%
\label{fig:mzi}%
\end{figure}

Light entering a MZI in TMSV state exits a lossless interferometer in the state
$\left\vert \psi_{\text{f}}\right\rangle=\hat{U}_{\text{MZI}}\left\vert \psi_{\bar{n}}\right\rangle$, where the MZI is described by the unitary transformation $\hat{U}_{\text{MZI}}$ (Fig. \ref{fig:mzi}).
This transformation, in terms of the field operators for the optical 
modes $\hat{a}$ and $\hat{b}$, is $
\hat{U}_{\text{MZI}}=\hat{U}\hat{P}_{\varphi} \hat{U}%
=\text{exp}\left(  \varphi\left(  \hat{a}^{\dagger}\hat{b}-\hat{b}^{\dagger
}\hat{a}\right)  /2\right) $, where $\hat{P}_{\varphi}=$exp$\left(  -i\varphi\hat{G}\right)  $ 
describes accumulation of a phase difference $\varphi$; and $\hat{U}=$exp$\left(  i\frac{\pi}{4}\left(  \hat{a}^{\dagger
}\hat{b}+\hat{a}\hat{b}^{\dagger}\right)  \right)  $ 
describes the 50-50 beam splitter, with a $\pi/2$ phase
shift for the reflected mode. In a linear 
medium the generator of phase evolution is  $\hat{G}=\left(  \hat{n}%
_{A}-\hat{n}_{B} \right)  /2$, where $ \hat{n}%
_{A}=\hat{a}^{\dagger}\hat{a}$ and $\hat{n}_{B}=\hat{b}^{\dagger}\hat{b}$
are the photon number operators in each mode.

Phase estimation is based on the detection of light at 
the outputs of MZI. Not all detection schemes are capable of exploiting 
the full 
potential of non-classical light to be super-sensitive and super-resolving. 
For example, intensity difference measurement, which is standard for optical
interferometry with coherent light, is not phase sensitive at all if TMSV
input is used \cite{ISI:000073584500105}. In our work, we consider parity 
detection for our measuring
scheme. The parity operator on output
mode $A$ is $\hat{\Pi}_{A}=\exp\left(i\pi\hat{n}_{A}\right)$. 
Parity was originally discussed
in the context of trapped ions by Bollinger \emph{et al.}
\cite{PhysRevA.54.R4649} and later adopted for optical interferometry by Gerry
and Campos \cite{PhysRevA.61.043811,PhysRevA.64.063814}. Super-sensitivity with
this detection
strategy has been shown for several classes of input states \cite{chiruvelli-2009}.
Finally, parity detection was also shown to allow better than classical resolution 
with coherent light at the shot-noise limit (SNL) \cite{gao-2009}.

Parity measurement on mode $A$ at the output of MZI is computed by
$\left\langle
\hat{\Pi}_{A}\right\rangle =\left\langle \psi_{\text{f}}\right\vert \hat{\Pi
}_{A}\left\vert \psi _{\text{f}}\right\rangle$. It has been shown in Ref. \cite{gao-2009} that the parity measurement on mode $A$ after the
final beam splitter is equivalent to the
measurement that is constructed from all the $\left\vert M,M^{\prime}\right\rangle\to\left\vert M^{\prime},M\right\rangle$ projectors 
as follows \cite{gao-2009}
\begin{equation}
\hat{\mu}_{AB}=\sum_{N=0}^{\infty}\sum_{M=0}^{N}\left\vert N-M,M\right\rangle
\left\langle M,N-M\right\vert,
\end{equation} 
acting on the inner modes of MZI, such that $\left\vert \psi\right\rangle =\hat{P}_{\varphi}\hat{U}\left\vert \psi_{\bar{n}}\right\rangle$. 
Our use of the $\hat{\mu}_{AB}$ operator here highlights the fact that parity detection combined with 50-50 beam
splitter provides a measurement scheme that includes all of the phase-carrying
off-diagonal terms in the two-mode density matrix \cite{gao-2009}.
Calculation of $\left\langle \hat{\mu}_{AB}\right\rangle$ simplifies significantly once it is noted that such an operator,
as well as a beam splitter, does not change the total number of photons in the 
state. Thus a
lossless MZI with the parity detection scheme does not mix twin Fock states with 
different number of photons giving:
\begin{equation}
\left\langle\hat{\Pi}_{A}\right\rangle=(1-t_{\bar{n}})\sum_{n=0}^{\infty}t_{\bar{n}}^n\left\langle\hat{\Pi}_{A}\right\rangle_n, \label{eq:piaver}
\end{equation}
where
$\left\langle\hat{\Pi}_{A}\right\rangle_n=\langle n,n\vert \hat{U}_{\text{MZI}}^{\dagger}\hat{\Pi}_{A}\hat{U}_{\text{MZI}}\left\vert n,n\right\rangle$ is the expectation value of the parity operator for twin Fock state input. 
In turn, the expression $\left\langle\hat{\Pi}_{A}\right\rangle_n=(-1)^nP_n\left(\cos\left(2\varphi\right)\right)$, given in terms of Legendre polynomials $P_n$,
could be found in Ref. \cite{PhysRevA.68.023810}.
Finally, one can identify our expression in Eq. (\ref{eq:piaver}) with the generating function for
Legendre polynomials \cite{mathbook} and arrive to the following:
\begin{equation}
\left\langle
\hat{\Pi}_{A}\right\rangle_{\varphi+\pi/2} =\left\langle \hat{\mu}_{AB}\right\rangle_{\varphi} =\frac{1}{\sqrt{1+\bar{n}(\bar{n}+2)\sin^2\varphi}},\label{eq:muab}%
\end{equation}
where an additional $\pi/2$ phase shift was introduced.
Eq.~(\ref{eq:muab}) is the central result of this paper and, in what follows, 
it will be used to study the resolution and sensitivity of our proposed
scheme.

\begin{figure}[ptb]
\includegraphics[width=8cm]{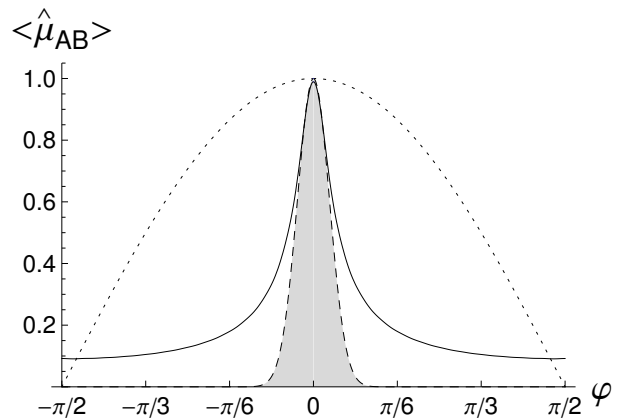}\caption{Measured signals at the output of MZI
with coherent light ($\bar{n}=100$) and TMSV ($\bar{n}=10$) inputs against accumulated phase difference.
Solid and dashed lines are the outputs of parity measurement for TMSV and coherent light, respectively.
Dotted line, given for comparison, is a scaled down output of intensity difference measurement on
the output of MZI fed with coherent light. TMSV with much smaller photon number performs as well as coherent
light.}%
\label{fig:comparison}%
\end{figure}

Let us compare here the signal outcomes of the TMSV scheme with $\bar{n}=10$ to
coherent-state-based optical interferometry with
$\bar{n}=100$ (see Fig. \ref{fig:comparison}).
Intensity difference measurement, with coherent state at the input of MZI, 
exhibits classical interference ---
a sinusoidal dependence on the phase with an intensity independent period of
$2\pi$. In the case of parity detection with coherent state input, it was shown in
Ref. \cite{gao-2009} that $\left\langle \hat{\Pi}_{A}\right\rangle
=$exp$\left(  -2\bar{n}\text{ sin}^{2}\left(  \varphi/2\right)  \right)  $
with a $2\pi$ period and a feature at the phase origin that is narrower than the
classical curve by a factor of $\delta\varphi=1/\sqrt{\bar{n}}$. In our case, the width of the feature
is further reduced by $\sqrt{\bar{n}+2}$ times. Therefore, the peak in Fig. \ref{fig:comparison}
is as narrow for a $\bar{n}=10$ TMSV as for a $\bar{n}=100$ coherent state input.

The other aim of optical interferometry is to minimize uncertainty, $\Delta\varphi$, of
the measured phase. The lowest bound on the uncertainty is inversely proportional to quantum Fisher
information $F_{Q}$ \cite{PhysRevLett.72.3439}. In the case
of a pure state $F_{Q}=4\Delta\hat{G}^{2}$, where $\Delta\hat{G}^{2}$ is
a variance of the phase evolution generator given above. Thus,%
\begin{equation}
\Delta\varphi_{\text{min}}^{2}=\frac{1}{4\Delta\hat{G}^{2}},
\label{eq:QFishlimit}%
\end{equation}
which depends on the state of the light used but not on the measurement.
It is in this spirit that we replaced coherent light with two-mode squeezed 
vacuum
in order to beat the sensitivity obtained in Ref. \cite{gao-2009}. 
Our analysis shows  that Eq. (\ref{eq:QFishlimit}), in the case of coherent 
light, limits
the attainable sensitivity to $\Delta\varphi_{\text{min}}^{2}=\bar{n}^{-1}$, shot noise, while for TMSV it sets 
much lower limit
$\Delta
\varphi_{\text{min}}^{2}=\left(\bar{n}(\bar{n}+2)\right)  ^{-1}<\bar
{n}^{-2}$. This means that TMSV has a potential
for super sensitive phase estimation, which has phase uncertainty better than
$1/\bar{n}$ and that is thus sub-Heisenberg. However, it remains to be seen whether 
sub-Heisenberg sensitivity could be obtained with a particular measurement,
namely parity.

\begin{figure}[ptb]
\includegraphics[width=8cm]{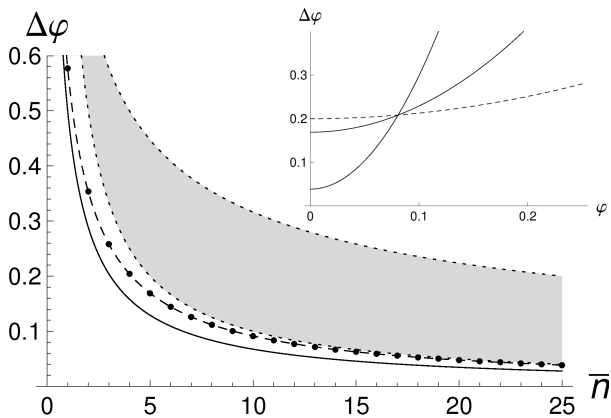}
\caption{Sensitivity of phase estimation obtained with parity measurement
at $\varphi=0$ (dashed) against average total 
photon number. Dotes are sensitivity estimation based on quantum Fisher 
information for integer values of $\bar{n}$. Shaded area is between dotted 
lines $1/\bar{n}$ and $1/\sqrt{\bar{n}}$. Solid line is for the Hofmann
limit discussed in the text. 
Inset: Sensitivity against actual values of accumulated phase difference. 
Solid lines for TMSV with $\bar{n}=5$ and $\bar{n}=25$;
dashed line for coherent light with $\bar{n}=25$.}%
\label{fig:SComp}%
\end{figure}

The variance of the phase estimation based on the outcome of the parity
measurement can be estimated as%
\begin{equation}
\Delta\varphi^{2}=\frac{1-\left\langle \hat{\mu}_{AB}\right\rangle ^{2}%
}{\left(  \partial\left\langle \hat{\mu}_{AB}\right\rangle /\partial
\varphi\right)  ^{2}}, \label{eq:sensdef}%
\end{equation}
which is a ratio of detection noise to the rate at which signal changes as function of
phase. 
We have shown that the rate of the signal change is much higher than in the
case of coherent state input. Thus, if parity measurement on the squeezed
vacuum is no noisier than on the coherent state, sensitivity improvement is expected.

The sensitivity of the phase estimation in our scheme can be estimated based
on Eq.~(\ref{eq:sensdef}) combined with Eq.~(\ref{eq:muab}):
\begin{equation}
\Delta\varphi=\frac{1+\bar{n}(\bar{n}+2)\sin^2\varphi}
{\vert\cos\varphi\vert\sqrt{\bar{n}(\bar{n}+2)}}, \label{eq:sens}%
\end{equation}
which is presented in Fig. \ref{fig:SComp} for the case of $\varphi=0$. 
It is clear that in this case sub-Heisenberg sensitivity
is obtained and that the lower bound defined by the quantum Fisher 
information is actually reached.
Expression in Eq. (\ref{eq:sens}) gives dependence of the phase sensitivity of our scheme 
on the actual phase difference as well:
\begin{equation}
\Delta\varphi\approx\frac{1}{\sqrt{\bar{n}(\bar{n}+2)}}\left(  1+\left(2\bar{n}(\bar{n}+2)+1\right)
\frac{\varphi^{2}}{2}\right), 
\end{equation}
where expansion near the phase origin was made.
This dependence is, in turn, compared to the one for a coherent state 
input, which
has the following functional dependence on the phase in the
vicinity of phase origin%
\begin{equation}
\Delta\varphi\approx\frac{1}{\sqrt{\bar{n}}}\left(  1+\left(  2\bar{n}+1\right)
\frac{\varphi^{2}}{8}\right)  .
\end{equation}
Dependence of the phase sensitivity in both cases is presented in the inset of 
Fig. \ref{fig:SComp} for $\bar{n}=5$ and $\bar{n}=25$ TMSV and for $\bar{n}=25$ coherent
state inputs.
Comparison shows that our scheme has superior sensitivity in the vicinity of phase origin
but degrades rapidly as actual phase difference deviates from zero.

Better than $1/\bar{n}$ phase sensitivity, demonstrated 
here by a linear MZI with parity detection and TMSV, is in 
violation of the so-called Heisenberg limit (HL). In what follows, we
will argue why HL is not the true limit, if HL is defined as $\Delta \varphi_{HL}=1/\bar{n}$. 

The traditional argument for the limit on the sensitivity of
the phase measurement comes from the Heisenberg uncertainty principle
for the phase and photon number $\Delta \varphi \Delta n\ge 1$. This argument is usually combined
with intuitive thinking that in the case of finite energy $\Delta n$ should 
be bounded by $\bar{n}$, that is $\bar{n}\ge \Delta n$. This argument is valid as long as a fixed 
photon number is assumed $\bar{n}=N$, but has to be treated
with caution for fluctuating photon numbers.
Such notions about the Heisenberg limit can be traced back to, for example,
work by Ou \cite{PhysRevA.55.2598} where he speculates that the fundamental
limit set by quantum mechanics on sensitivity is 
the Heisenberg 
limit $\Delta\varphi \ge 1/\bar{n}$ since all analysis up until now had not shown better than $1/\bar{n}$ 
sensitivity. Ou's conjecture is in fact not supported by his cited 1986 paper by Yurke \emph{et al.}, 
on SU(1,1) interferometers \cite{PhysRevA.33.4033},
where Eq. (9.31) was published with a typographical error --- there is a minus sign where there should be a plus.

Recent progress in quantum metrology has demonstrated the importance of photon number 
fluctuations for phase estimation. In order to better account for photon number fluctuations,
Hofmann in Ref. \cite{hofmann:033822} suggested a more direct definition of the ultimate 
quantum limit of phase sensitivity $\Delta\varphi^2 \ge 1/\langle \hat{n}^2\rangle$, where $\langle \hat{n}^2\rangle$ indicates averaging over the 
squared photon numbers. Thus, in the case of high photon number fluctuations,
$\Delta n^2=\langle \hat{n}^2\rangle-\langle \hat{n}\rangle^2>0$,
the Hofmann limit allows for better sensitivity of the phase measurement than the
Heisenberg limit. Clearly $\langle \hat{n}^2\rangle$ contains direct information
about fluctuations where $\langle \hat{n}\rangle^2$ does not.

\begin{figure}[ptb]
\includegraphics[width=8cm]{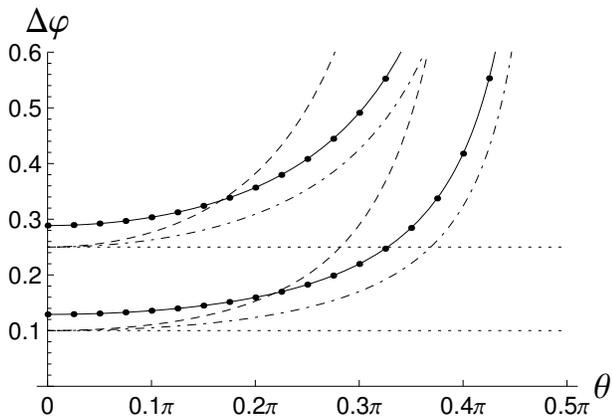}\caption{Phase estimation sensitivity $\Delta\varphi$ for the state 
$\hat{\rho}(n,\theta)$ with $n=2$ (top four) and $n=5$ (bottom four) against $\theta$.
Solid lines represent sensitivity with parity measurement which saturates limit set by quantum Fisher information 
(dots for a few selected values of $\theta$). Dashed lines represent HL sensitivity estimation based on averaged total 
photon number: $1/\bar{n}$. Dot-dashed lines give Hofmann's estimation of sensitivity discussed in the text.}%
\label{fig:SensTFV}%
\end{figure}

In the case of our TMSV with parity, sensitivity of phase estimation is better than allowed by the Heisenberg limit,
although marginally, but it is never better than $1/\sqrt{2\bar{n}^2+2\bar{n}}$, which is the Hofmann limit. 
It is also never below the quantum Cramer-Rao lower bound set by the quantum 
Fisher information of the state at an area of phase accumulation.

In order to demonstrate that the maximal phase sensitivity could be underestimated by the Heisenberg limit
if photon number fluctuations are neglected, consider the following state
$\hat{\rho}(n,\theta) =\text{sin}^2\theta\left\vert
0,0\right\rangle\langle0,0\vert +\text{cos}^2\theta\left\vert n,n\right\rangle\langle n,n\vert$, which has $\bar{n}=2n\cos^2\theta$.
This state could appear in the context of a probabilistic twin Fock state generation
with parity detection on a single output mode, since such a detection would not
distinguish vacuum contribution from the twin Fock contribution when all
photons were routed out in the other port.

Based on quantum Fisher information, this state is capable of providing sensitivity
$\Delta\varphi^{2}=1/\left(2 n\left(n+1\right)\text{cos}^{2}\theta\right)$, which is obtainable by parity measurement.
This dependence of the phase sensitivity is presented in Fig. \ref{fig:SensTFV}
for $n=2$ and $n=5$,
where the presence of the vacuum, $\theta>0$, degrades the sensitivity but 
allows for $1/\bar{n}$ Heisenberg limit to be beat! However, the Hofmann limit $\Delta\varphi^{2}=1/\left(4 n^2\text{cos}^{2}\theta\right)$
tracks the phase sensitivity well; without being beaten!

There does exist another limit based on
the highest number of photons in the state --- $1/N$, with $N=2n$ for the state considered here.
However, it is not as useful as the Hofmann limit for a number of reasons:
a) it overestimates the sensitivity as it does for $\hat{\rho}(n,\theta)$;
b) information about $N$ is not readily available in experiments;
c) for states, such as coherent and squeezed vacuum, $N=\infty$.

Finally, implementation of parity detection needs to be discussed. 
In proof of principle experiments, a highly efficient photon number-resolving detector
could be used. Such detectors with 95\% efficiency and number resolving capabilities in 
the tens of photons have been demonstrated \cite{ISI:000184336600067,ISI:000254121300021,wildfeuer:043822,khoury:203601}. 
However, for more practical applications, knowledge about exact photon numbers
is excessive.
We conjecture that a scheme, which does not require photon counting, exists, perhaps through the
exploitation of optical non-linearities \cite{PhysRevA.72.053818}, 
or projective measurements, and this is an area of ongoing research.

In conclusion, the main result of this paper is our demonstration that optical interferometry
with two-mode squeezed vacuum and parity detection provides an interferometric
metrology strategy with sensitivity $\Delta\varphi<\bar{n}^{-1}$ and
resolution $\bar{n}^{-1}$ times better than the resolution of standard (classical) interference.

\begin{acknowledgments}
We would like to acknowledge support from the Army Research Office, the Boeing
Corporation, the Department of Energy, the Foundational
Questions Institute, the Intelligence Advanced Research Projects Activity, and
the Northrop-Grumman Corporation.
\end{acknowledgments}


\end{document}